\documentstyle[aps,prd,preprint,eqsecnum,epsf]{revtex}
\input epsf.tex

\newcommand{\be}{\begin{equation}}
\newcommand{\ee}{\end{equation}}
\newcommand{\ba}{\begin{eqnarray}}
\newcommand{\ea}{\end{eqnarray}}
\newcommand{\bas}{\begin{eqnarray*}}
\newcommand{\eas}{\end{eqnarray*}}

\newcommand{\lbr}{\left <}
\newcommand{\rbr}{\right >}

\tightenlines

\begin{document}

\draft

\title{Holography for degenerate boundaries}
\author{Marika Taylor-Robinson
\thanks{email: M.M.Taylor-Robinson@damtp.cam.ac.uk}}
\address{Department of Applied Mathematics and
      Theoretical Physics, \\ Centre for Mathematical Sciences,
      \\ Wilberforce Road, Cambridge CB3 0WA, United Kingdom }
\date{\today}

\maketitle

\begin{abstract}
{We discuss the AdS/CFT correspondence for negative curvature Einstein
  manifolds whose conformal boundary is degenerate in the sense that
  it is of codimension greater than one. In such manifolds, 
  hypersurfaces of constant radius do not blow up 
  uniformly as one increases the radius; examples 
  include products of hyperbolic spaces and the Bergman metric. We
  find that there is a well-defined correspondence between the IR
  regulated bulk theory and conformal field theory defined 
  in a background whose degenerate geometry is regulated by the 
  same parameter. We are hence able to make sense of supergravity in
  backgrounds such as $AdS_{3} \times H^{2}$. 
}
\end{abstract}


\section{Introduction}
\label{sec:intro}

Following the now famous conjecture of Maldacena \cite{Ma},
\cite{GKP}, \cite{W1}, \cite{A} relating supergravity on anti-de Sitter
spacetimes to a conformal field theory in one less dimension there has
been a great deal of interest in supergravity compactifications on
anti-de Sitter spaces.
One of the most interesting features of the AdS/CFT correspondence is
that it provides an example of the holographic principle \cite{Su}. In
the context of the Maldacena conjecture, holography was first
discussed in detail in \cite{SW} whilst discussions of holography in
cosmology have appeared in \cite{FS}, \cite{Bou}.

However, the holographic description of the anti-de Sitter bulk theory
relies heavily on special features of the boundary, namely
that the conformal boundary in the sense of Penrose \cite{Pen} is a
non-degenerate manifold of codimension one.  
Although most of the physical negative curvature solutions
that one is interested in considering, such as black holes, 
satisfy these properties, several classes of negatively
curved spacetimes do not. 

In particular, coset spaces such as $SO(3,1)/SO(3)$,
Bergman type metrics and products of hyperbolic spaces have conformal
boundaries of codimension greater than one, which we will refer to as
degenerate boundaries. Typically, if $1/\epsilon$ characterizes the radius of
the $(d+1)$-dimensional spacetime, then the 
volume of a hypersurface of constant $\epsilon$ behaves as $\epsilon^{\alpha}$ 
with $\alpha > - d$; the induced hypersurface does not blow up
uniformly as $\epsilon \rightarrow 0$. 
First steps towards understanding the bulk boundary correspondence in
these cases were taken in \cite{BSV} and the purpose of this paper is
to take this correspondence further and deal with some of the
unresolved issues of \cite{BSV}.

In particular, although the authors of \cite{BSV} were able to analyse
the correspondence between scalar fields in the bulk and boundary 
scalar operators for the Bergman metric, they did not find such an
interpretation for coset spaces which we shall do here. Furthermore,
to properly formulate the bulk/boundary correspondence, one needs to
understand how the bulk partition function gives rise to the partition
function for the boundary conformal field theory. 

When the conformal boundary is non-degenerate, there is a well
understood way of defining the bulk (Euclidean) action without
introducing a background \cite{BK}, \cite{BK2}, \cite{KLS}. 
One introduces local boundary counterterms
into the action, which remove all divergences and leave a finite
action corresponding precisely to the partition function of the
conformal field theory. 

This procedure must break down when the boundary is degenerate, since
there is no conformal frame in which the hypersurface 
radius $\epsilon$ does not appear in the metric. 
This means that one cannot hope
to eliminate all radius dependence from the partition function; put
differently, as one takes the limit $\epsilon \rightarrow 0$, the
partition function for the conformal field theory must diverge, since
the geometry is becoming highly degenerate. 

So the question is whether one can make sense of a correspondence
between bulk and boundary partition functions. We will show that one
can, provided that one takes the correspondence to be at finite
radius. That is, an IR cutoff in the bulk theory will appear as a
regulation of the conformal field theory target geometry. This is a novel
manifestation of the IR/UV correspondence \cite{SW}. 

In several recent papers the relationship between the Randall/Sundrum scenario
\cite{RS} and AdS/CFT correspondence has been explored \cite{G}, \cite{V},
\cite{W2}. In this context, one can view the five-dimensional
negative curvature spacetime to be a construction from a symmetric
non-degenerate four-dimensional brane world. In \cite{RS}, \cite{G}, the brane
world lives at finite radius $R$, and the induced action on the brane
includes Einstein gravity plus corrections. For our ``degenerate''
brane worlds, however, the induced action on the brane does not
include Einstein gravity, even when the brane is four-dimensional. 

\bigskip

The plan of this paper is as follows. In \S \ref{two}, we discuss
classes of metrics which have degenerate conformal boundaries. In \S
\ref{three}, we discuss the interpretation of the gravitational bulk
action for such spacetimes in terms of a dual conformal field
theory. In \S \ref{four}, we analyse how correlation functions for
scalar operators in the conformal field theory may be derived 
from bulk massive scalar fields.

\section{Classes of metrics with degenerate boundaries} \label{two}
\noindent

Suppose that $M$ is a complete Einstein manifold of negative curvature
and dimension $d+1$ which has a conformal boundary which a
$d$-manifold $N$. This means that the metric of $M$ can be written
near the boundary as 
\be
ds^2 = \frac{du^2}{u^2} + \frac{1}{u^2} g_{ij}(u,x) dx^i dx^j,
\ee
where $u$ is a smooth function with a first order zero on $N$ which is
positive on $M$. Usually we assume that $g_{0} = g(0,x)$ is a 
non-degenerate metric on $N$, independently of how we take the limit
$u \rightarrow 0$. However, more general negative curvature manifolds
can have conformal boundaries which are degenerate in the sense that 
$g(\epsilon,x)$ is divergent as $\epsilon \rightarrow 0$. The
vielbeins for such a metric will not all be finite as $\epsilon
\rightarrow 0$; at least one will tend to zero or diverge. 

Let us refer to $N_{\epsilon}$ as the regulated boundary; then
$N_{\epsilon}$ will have a natural conformal structure. We will show
in the following section that $N_{\epsilon}$ must have negative
curvature of the order of the $(d+1)$-dimensional cosmological
constant when $N$ is degenerate. 
We expect there to be a correspondence between conformal
field theory on $N_{\epsilon}$ with a UV cutoff $1/\epsilon$ and
quantum gravity on $M$ with an IR cutoff $\epsilon$. To investigate
this correspondence we will as usual consider the relation between bulk
and boundary partition functions. Before we do so, however, it will be
helpful to give several examples of spaces which have degenerate
boundaries; we will be using these explicit examples to illustrate
general arguments in the following sections. 
 
\bigskip

We will consider here two generic classes of negative curvature manifolds
which have degenerate conformal boundaries. Let us normalise 
the $(d+1)$-dimensional Einstein action such that
\be
I_{\rm{bulk}} = - \frac{1}{16 \pi G_{d+1}} \int_{M} d^{d+1}x \sqrt{g}
( R_{g} + d(d-1) l^2 ) - \frac{1}{8 \pi G_{d+1}} \int_{N} d^dx
\sqrt{\gamma} K, \label{action}
\ee
where $R_{g}$ is the Ricci scalar and $K$ is the trace of the
extrinsic curvature of the boundary $N$ embedded in $M$. 

An example of the first class of degenerate boundary manifolds
was considered in \cite{BSV}; it is included in 
the family of Einstein-K\"{a}hler metrics of the form
\ba
ds^2 &=& (l^2 r^2 + \frac{1}{4} - \frac{k}{r^4})^{-1} dr^2 + r^2 
(l^2 r^2 + \frac{1}{4} - \frac{k}{r^4})(d\psi + 2 \cos\theta d\phi)^2
\\ && \hspace{85mm} + r^2
(d\theta^2 + \sin^2\theta d\phi^2), \nonumber
\ea
which was constructed in \cite{PP}. 
Regularity requires that the periodicity of $\psi$ is
\be
\beta = \frac{8 \pi}{k} = \frac{4 \pi}{U'(r_{+}) r_{+}},
\ee
where the horizon location is $r_{+}$ and $U(r)$ is the metric
function. The boundary admits the conformal structure
\be
ds^2 = (l^2 r^2 + \frac{1}{4} - \frac{k}{r^4}) (d\psi + 2 \cos\theta
d\phi)^2 + (d\theta^2 + \sin^2\theta d\phi^2),
\ee
which is a squashed three-sphere of the form
\be
ds^2 = l_{3}^2 (d\tilde{\psi} + \cos\theta d\phi)^2 + (d\theta^2 +
\sin^2\theta d\phi^2),
\ee
and becomes degenerate as we take the limit $r \rightarrow
\infty$. The $k = 0$ metric is the Bergman metric which is the group
manifold $SU(2,1)$ discussed
in \cite{BSV}; it will be convenient to use here an alternative 
form of the Bergman metric 
\be
ds^2 = \frac{2}{l^2}
[d\rho^2 + \frac{1}{4} \sinh^2 \rho \cosh^2 \rho (d\psi + \cos
\theta d \phi)^2 + \frac{1}{4} \sinh^2 \rho (d\theta^2 + \sin^2 \theta
d\phi^2)],
\ee
where the prefactor ensures that the curvature behaves as $R_g = - 12 l^2$.
Although the boundary of the Bergman metric becomes degenerate in the
infinite limit, there is a very natural choice of conformal boundary since
this space, like others in the same family, 
is constructed by radially extending a $U(1)$ bundle
over an Einstein space. 

Much that one says about the Bergman metric can
also be extended to other radial extensions of bundles over compact
spaces. Such spaces have been considered in the past 
in the context of compactifications \cite{DNP} and a
number of other examples are known. For example, in eight dimensions
one could consider an Einstein hyper-K\"{a}hler manifold of the form
\be
ds^2 = d\rho^2 + \frac{1}{4} \sinh^2 \rho [ d\theta^2 + \frac{1}{4}
\sin^2 \theta \omega_{i}^2 + \frac{1}{4} \cosh^2 \rho ( \nu_i + \cos
\theta \omega_i)^2],
\ee
where $\nu_{i}, \omega_i$ are left-invariant forms satisfying $SU(2)$
algebras. This metric is obtained by analytically continuing the
standard Fubini-Study Einstein metric on $HP^2$, the quaternionic
projective plane. Hypersurfaces of constant $\rho$ are squashed
seven-spheres which become infinitely squashed as one takes the
boundary to infinity. Analysis of the AdS/CFT correspondence in this
case would be very similar to that for the Bergman metric. 
One would also expect that one could find an Einstein metric which is a
radial extension of a twisted $S^2$ bundle over $S^2$. Hypersurfaces
of constant radius should correspond to off-shell extensions of the
Page metric \cite{P}. 

We should mention that these metrics do not have a nice physical Lorentzian
interpretation. The existence of a topologically non-trivial degenerate 
conformal boundary is related to the non-trivial fibration of $U(1)$
coordinates which would have to be analytically continued to give a
Lorentzian metric. Analytically continuing the Bergman metric leads to
a Lorentzian metric which is complex and whose physical interpretation
is unclear. Furthermore, not only are these metrics not supersymmetric
(the Bergman metric is certainly not supersymmetric since there is no
supergroup which has $SU(2,1)$ as its bosonic part) but also many such
metrics will not even admit a spin structure. 

Given the problems with the Lorentzian interpretation, we shouldn't be
surprised if the thermodynamic quantities in the conformal field
theory take unphysical values. It is well known \cite{CH} that a 
generic quantum field theory within a causality violating background
will exhibit pathologies such as negative entropy which one interprets
as reflecting the unphysical nature of the background and we will see
similar phenomena here.  
However, although such backgrounds are not of great physical interest
in string theory they provide interesting examples of
a more general AdS/CFT correspondence. There may be other backgrounds 
in string theory which are physically interesting and 
exhibit similar degenerate behaviour. 

\bigskip

The second category of manifolds in which we are interested is
characterized by the existence of more than one ``radial'' coordinate.
The simplest example which we will consider in the most detail is
the product of hyperbolic manifolds of dimensions $d_1$ and $d_2$
respectively:
\be
ds^2 = \frac{(d_1-1)}{(d_1 + d_2 -1) l^2 z_1^2} (dz_1^2 + d{\bf{x}}_1^2) +  
\frac{(d_2 - 1)}{ (d_1 + d_2 - 1) l^2 z_2^2} (dz_2^2 + d{\bf{x}}_2^2).
\ee
``Infinity'' in this metric is obtained by taking either
$z_1$ or $z_2$ to zero; however, for a holographic principle to be
formulated we need to define a boundary of codimension one. One way
to do this would be to effectively divide the boundary into two parts:
consider a hypersurface of constant $z_1 = \epsilon_1 << 1$ and a 
hypersurface of constant $z_2 = \epsilon_2 << 1$ which are glued
together along the hypersurface of codimension two $z_1 = \epsilon_1;
z_2 = \epsilon_2$. However, since one would need to be careful about
boundary conditions for fields along this join it is easier to work
with the conformally equivalent surface defined by setting 
\be
z_1 = u \cos \theta, z_2 = u \sin \theta,
\ee
so that for example when $d_1 = d_2 = 2$ the metric becomes
\ba
ds^2 &=& \frac{1}{3 l ^2} \lbrace \frac{2 du^2}{u^2} + \frac{2 du d\theta}{u} (
\frac{\cos\theta}{\sin\theta} - \frac{\sin\theta}{\cos\theta}) + 
 \frac{(\sin^4 \theta + \cos^4 \theta)}{\sin^2\theta \cos^2 \theta}
d\theta^2 \\
&& \hspace{70mm} + \frac{d{x}_1^2}{u^2 \cos^2 \theta} + \frac{d{x}_2^2}{u^2
  \sin^2 \theta} \rbrace. \nonumber
\ea
With this choice of coordinates the induced metric on constant $u$
hypersurfaces is
\be
ds^2 = \frac{d\theta^2}{2 \sin^2\theta \cos^2 \theta} +
\frac{d{x}_1^2}{u^2 \cos^2 \theta} + \frac{d{x}_2^2}{u^2 \sin^2 
\theta}, \label{u1}
\ee
which is degenerate as $u \rightarrow 0$ and further degenerates when
$\theta \rightarrow 0, \pi/2$. Notice that the metric is non-singular
in this conformal frame, with the limits in $\theta$ corresponding to
non-singular tubes in the Euclidean metric.

\bigskip

As well as products of negative curvature manifolds, various coset
spaces exhibit a similar degenerate behaviour. In particular, one can
analytically continue manifolds which were considered in the context
of supergravity compactifications, reviewed in \cite{DNP}. The coset
space $SO(3,1)/SO(2)$ possesses an Einstein metric
\be
ds^2 = \frac{1}{9} (d\psi + i \cosh\rho_1 d\phi_1 + i \cosh \rho_2 d
\phi_2)^2 + \frac{1}{6} \sum_i (d\rho_i^2 + \sinh^2 \rho_i d\phi_i^2),
\label{coset}
\ee
where we have taken $l^2 = \frac{1}{4}$. 
The Euclidean metric is not real, but one can find a Lorentzian
section (discussed
in \cite{BSV}) which is; however, our Lorentzian theory will still 
exhibit pathologies since there are closed timelike curves. 

The most useful choice of boundary in this case is probably to take 
\be
e^{\rho_1} = \frac{1}{u \cos \theta}; \hspace{5mm} e^{\rho_2} =
\frac{1}{u \sin \theta},
\ee
in analogy to the above. Then as $u \rightarrow 0$ the leading order
induced metric on a constant $u$ surface is
\ba
ds^2 &=& \frac{d\theta^2}{12 \sin^2\theta \cos^2 \theta} + 
\frac{d{\phi}_1^2}{72 u^2 \cos^2 \theta} + \frac{d{\phi}_2^2}{72 u^2 \sin^2 
\theta} + \frac{1}{9} d\psi^2 \label{cobound} \\ 
&& \hspace{30mm} + \frac{i}{9u} d\psi (\frac{d\phi_1}
{\cos \theta} + \frac{d\phi_2}{ \sin \theta}) - \frac{1}{18 u^2 \sin
  \theta \cos \theta} d\phi_1 d \phi_2 , \nonumber
\ea
which has a $so(2)^3$ symmetry group and is degenerate as $\theta
\rightarrow 0, \pi/2$. 

\bigskip

All of the examples so far are symmetric negative curvature
manifolds with degenerate boundaries. One could also start with a
symmetric manifold which is degenerate as some parameter $u$ is taken
to zero. One would then re-interpret $u$ as an effective radius
and radially extend the hypersurface to construct an Einstein manifold
in one higher dimension. This point of view was discussed in
\cite{KSo} in the context of negative curvature spacetimes with 
non-degenerate boundaries. It will become clearer in the following
section how one would radially extend the hypersurface in the
degenerate case. 

The conformal symmetry group will be larger than in most 
of the examples given above, and so one will be able to fix more
quantities in the conformal field theory. However, calculating bulk
quantities will be correspondingly more difficult, since the bulk
symmetry group will in general be smaller. This will be particularly
relevant when trying to derive, for example, scalar correlation functions
from bulk actions for massive scalar fields. 

\section{Regularisation of the Euclidean action} \label{three}
\noindent

One of the most interesting developments arising from the AdS/CFT
correspondence has been the use of counterterms in the Euclidean action to
define the action independently of background for negative (and in
certain limits zero) curvature manifolds \cite{HK}, \cite{BK}, \cite{KLS}. 
Consider an Einstein manifold which satisfies the field equations
derived from (\ref{action}), whose metric near the boundary can be
written in the form
\be
ds^2 = \frac{dx^2}{l^2 x^2} + \frac{1}{x^2} \gamma_{ij} dx^i
dx^j, \label{ab} 
\ee
where $\gamma$ is finite and non-degenerate on the boundary itself.
In this section we will consider only four-dimensional spacetimes,
since this is the dimensionality of the explicit examples which we
will use. Following a theorem by Fefferman and Graham \cite{FG}, \cite{GL} 
the conformal metric $\gamma_{ij}$ can be written \cite{HK}, \cite{G} as
\be
\gamma_{ij} = \gamma^{0}_{ij} + x^2 \gamma_{ij}^{2} 
+ x^4 \gamma^{4}_{ij} +...,
\label{eqA}
\ee
where $\gamma^2$ is defined in terms of the curvature of $\gamma^{0}$
as \cite{KSo} 
\be
\gamma^2_{ij} = -\frac{1}{4 l^2} (R^{0}_{ij} - \frac{1}{4}
R^{0} \gamma_{ij}^{0}),
\ee
and $\gamma^{4}$ depends on fourth derivatives of $\gamma^0$.
$\gamma^0$ is independent of $x$ when the conformal boundary
is non-degenerate. Note that the choice of bulk conformal frame - or
in other words, the magnitude of the cosmological constant - combined
with the existence of a non-degenerate conformal boundary 
effectively fixes the coefficient of the $dx^2$ term in (\ref{ab})
\cite{FG} to be $1/l^2$. 
We emphasise this point since it will be important in what
follows. 

We can then formally expand the Einstein action as \cite{HK}
\ba
I_{\rm{bulk}} &=& - \frac{1}{16 \pi G_{4}} \int_{N_{\epsilon}} d^4x \sqrt{g} 
(R(g) + 6 l^{2}) - \frac{1}{8 \pi G_4} \int_{N_{\epsilon}} 
d^3x \sqrt{\gamma} K; \nonumber \\
&=& \frac{1}{16 \pi G_{4}} \int_{N_{\epsilon}} d^3x \sqrt{\gamma^{0}} 
( - \frac{4 l}{ \epsilon^3} +
\frac{16 l}{\epsilon} \rm{tr} (\gamma^{0})^{-1} \gamma^{2} + ...); \\
&=& - \frac{l}{4 \pi G_4} \int_{N_{\epsilon}} d^3x \sqrt{\tilde{\gamma}} (
1 + \frac{1}{4l} R(\tilde{\gamma}) + ... ), \nonumber
\ea
where the inverse radius of the hypersurface over which we integrate,
$\epsilon \ll 1$, is an IR regularisation parameter
and $x^2 \tilde{\gamma} = \gamma$. The ellipses indicate
non-divergent terms which we have omitted. 
The second equality is obtained by using
the expansion of the boundary metric (\ref{eqA}), (\ref{ab}) 
and integrating the bulk action explicitly. The key point is that
since there are 
only a finite number of divergent terms one can introduce a local counterterm 
action $I_{\rm{ct}}$ 
dependent only on $\tilde{\gamma}_{ij}$ and its covariant derivatives \cite{BK}
\be
I_{\rm{ct}} =  \frac{l}{4  \pi G_4} \int_{N_{\epsilon}} 
d^3x \sqrt{\tilde{\gamma}} (1 + \frac{1}{4l} R({\tilde{\gamma}})). \label{rsc}
\ee
Then the content of the AdS/CFT conjecture is that 
we make the identification that 
\be
I_{\rm{bulk}} = W_{\rm{cft}} + I_{\rm{ct}},
\ee
where we take the boundary to be the true conformal boundary
and $W_{\rm{cft}}$ is the (finite) partition function for the
conformal field theory. The purpose of this section is to show how and
why this analysis breaks down when the boundary becomes degenerate. 

\subsection{Degenerate boundaries and counterterm regularisation}
\noindent

For the manifolds with degenerate boundaries considered here,
the expansion of $\gamma_{ij}$ given in (\ref{eqA}) breaks down; 
its derivation in fact relies on the existence of a non-degenerate conformal
boundary of codimension one \cite{FG}, \cite{GL}. 
We are going to consider a more general form for the expansion of the  
metric near the conformal boundary 
\be
ds^2 = \frac{dx^2}{l^2 x^2} + x^{-\delta} \gamma_{ij} dx^{i} dx^j.
\label{cd}
\ee
We will assume that 
there is a well-defined expansion for the boundary metric of the form 
\be
\gamma_{ij} = \gamma^{0}_{ij} + \gamma^{2}_{ij} + ....,, \label{eqB}
\ee
but $\gamma^{0}$ will now depend on $x$. There is a
preferred frame in which its determinant is
independent of $x$: this choice of $\gamma^{0}$ is natural if one
requires that both the determinant and the inverse metric are
well-defined as $x \rightarrow 0$. We assume from here on that this
choice of normalisation for $\gamma^{0}$ is imposed in the metric
(\ref{cd}), which along with the normalisation of the coefficient of
$dx^2$ in (\ref{eqB}), effectively determines the 
choice of the coordinate $x$ given an Einstein metric satisfying
(\ref{action}). $\gamma^2$ will be subleading in the sense that its
determinant behaves as a positive power of $x$ as one takes the limit
$x \rightarrow 0$. As we will see later on in this section, the
explicit metrics given in \S \ref{two} all admit an expansion of this
form. 

Suppose
that $\gamma^{0}$ effectively degenerates to a $p$-dimensional metric 
in the limit that $x \rightarrow 0$, where $p$ will be determined by
the number of independent vielbeins. Note that the $p$-dimensional
metric does not in general have a non-zero determinant nor will the
associated vielbeins be closed. In fact, the Bergman metric
degenerates to a metric with zero determinant whose associated single
vielbein is not closed. Then a typical degenerate boundary metric 
 $\gamma^{0}$ might be written as
\be
\gamma^{0}_{ij} = x^{-1} h^{(p)}_{ij} + x^{\frac{p}{3-p}}
h^{(3-p)}_{ij}, \label{eqC}
\ee
where $h^{q}$ is of dimension $q$ in the sense defined above. 
Now $\gamma^{2}$ is defined by requiring that the
metric (\ref{cd}) satisfies the Einstein equations expanded out in
powers of $x$. 
By analysing the field equations, however, 
we find that although $\gamma^2$ is well-defined, it is not
a covariant tensor: it cannot be written in terms of $\gamma^0$ and
its curvature invariants. 
The definition of $\gamma^2$ for (\ref{eqC}) is not particularly 
illuminating since it is not generic; it
involves second derivatives of $h^{p}$ and $h^{3-p}$ as one would expect.

Using the asymptotic form for the metric, 
the Einstein action for a manifold with a degenerate boundary
(\ref{cd}) can then be written as   
\ba
I_{\rm{bulk}} &=& - \frac{1}{16 \pi G_{4}} \int_{N_{\epsilon}} d^4x \sqrt{g} 
(R(g) + 6 l^{2}) - \frac{1}{8 \pi G_4} \int_{N_{\epsilon}} 
d^3x \sqrt{\gamma} K, \nonumber \\
&=& \frac{l}{16 \pi G_4} \int_{N_{\epsilon}} d^3x \sqrt{\tilde{\gamma}}
(\frac{4}{\delta} - 3 \delta) + I_{\rm{nl}}, \label{surf}
\ea
where the second equality follows from explicitly substituting the
metric (\ref{cd}) into the action and integrating. $\tilde{\gamma} =
\epsilon^{-\delta} \gamma$ is the metric induced on a codimension one
hypersurface of constant $\epsilon$ and the integral is taken over a
hypersurface of constant $\epsilon \ll 1$. 

The second part of
the action, $I_{\rm{nl}}$, includes non-local terms and cannot be
expressed covariantly in terms of the boundary metric. This term in 
the action is not in general finite, but diverges as one takes the limit
$\epsilon \rightarrow 0$. However, the leading order divergence of the
bulk action (which behaves as $\epsilon^{-3\delta/2}$)
can be removed by subtracting the first term in (\ref{surf}); 
this follows from the condition that $\gamma^2$ is
subleading to $\gamma^0$. One will be left with a leading order
divergent term in $I_{\rm{nl}}$ 
which behaves as $\epsilon^{-a}$ with $a < 3 \delta /2$. Note that the
first term in the action (\ref{surf}) agrees with that for non-degenerate 
boundaries when one takes $\delta = 2$.

\bigskip

One implication of the above is that one
cannot introduce local counterterms to remove the divergence of the
bulk action as $\epsilon \rightarrow 0$. Suppose we tried to take a 
counterterm action of the form \cite{BK}
\be
I_{ct} = \frac{1}{16 \pi G_4} \int_{N_{\epsilon}} 
d^3 x \sqrt{\tilde{\gamma}} [ a_{0}[\delta]
+ a_{1} R(\tilde{\gamma}) + a_{2} (R(\tilde{\gamma})^2 + b_2
R^{ij}(\tilde{\gamma}) R_{ij}(\tilde{\gamma}) + ....).
\ee
Provided we pick the first coefficient
$a_{0}[\delta]$ according to (\ref{surf}) we can remove the leading
order divergence - but there is no generic way to define the other
coefficients. In fact, even choosing $a_{0}[\delta]$ in this way
really represents a fine-tuning which we are not allowed to do. 
One more general grounds, we can see that this series cannot be convergent
in $\epsilon$ without adjusting the coefficients to each solution.
Since the curvature invariants 
of hypersurfaces of constant $\epsilon$ are of order $l$ for
degenerate boundaries (compared to invariants of order $\epsilon$ 
to positive powers for non-degenerate boundaries), there is no small
expansion parameter and no reason for the series to converge. 

This behaviour of the curvature invariants follows 
from the Gauss-Codacci condition for the induced hypersurface
\be
R(\tilde{\gamma}) = (K^2 - K_{ab} K^{ab} - 6 l^{2}),
\ee
where $K_{ab}$ is the extrinsic curvature of the
hypersurface and $K$ is its trace as before. 
For a metric which can be written in the form (\ref{eqA}) 
with $\gamma^{0}$ non-degenerate, then
\be
K^2 = 9 l^2 + {\cal{O}}(\epsilon^2); \hspace{5mm} 
K_{ab} K^{ab} = 3 l^2 + {\cal{O}}(\epsilon^2),
\ee
and so the curvature invariants of the hypersurface behave as positive
powers of $\epsilon$, which is really the basis of the counterterm
subtraction procedure \cite{BK}.  
However, if $\gamma^{0}$ is degenerate and, for example, 
of effective dimension $p$, then
\be
K^2 = \frac{(3+p)^2}{4}l^2 + {\cal{O}}(\epsilon^2); \hspace{5mm} 
K_{ab} K^{ab} = \frac{3(p+1)}{4}l^2 + {\cal{O}}(\epsilon^2),
\ee
and so, as previously mentioned, 
the curvature of the hypersurface is {\it negative} and of 
order $l$. It is a generic feature of spaces with degenerate conformal
boundary that the induced metric on the boundary has finite negative
curvature, rather than an infinite curvature radius as is usual.
We should perhaps mention here that some of the analysis of
the AdS/CFT correspondence relies on non-negative curvature of the CFT
background spacetime. In particular, the discussion in \cite{WY}
relies specifically on a conformal boundary of positive scalar
curvature. Interesting issues that arise even in the non-degenerate
case when the boundary has negative curvature are discussed in
\cite{E}. 

\bigskip

Of course after a little reflection 
we should not be surprised that local counterterms cannot
remove the divergence of the bulk action. Since 
$\epsilon$ appears explicitly in the conformal field theory background
geometry, we cannot expect the partition function to be independent of
this parameter. Furthermore we should probably 
expect the partition function to diverge as the geometry becomes degenerate. 

As a simple example let us 
consider a generic conformal field theory on a $d$-dimensional background
\be
ds^2 = u^{d-1} d\tau^2 + u^{-1} h_{ij} dx^{i} dx^{j},
\ee
where $\tau$ is the trivially fibered imaginary time coordinate with
period $\beta$ and $h$ is a non-degenerate metric. We suppose that 
$u$ corresponds to a radial parameter in the bulk theory, and this
metric is conformal to that induced on hypersurfaces of constant $u$. 
As $u \rightarrow 0$, the metric will become degenerate, 
although in this (preferred) conformal frame 
the determinant remains regular. Suppose we now conformally rescale the
metric such that 
\be
\tilde{ds}^2 = d \tilde{\tau}^2 + h_{ij} dx^{i} dx^{j}, 
\ee
where we have defined a new imaginary time coordinate $\tilde{\tau} =
u^{\frac{d}{2}} \tau $. In this conformal frame it is trivial to write
down the main dependence of the partition function since the 
effective temperature is high in the degenerate limit: 
$\tilde{\beta} \rightarrow 0$ as $u \rightarrow 0$. This means that
the partition function for the conformal field theory behaves as
\be
W_{\rm{cft}} \sim T^{d-1} u^{-\frac{d(d-1)}{2}},
\ee
where $T$ is the inverse of $\beta$, and would be interpreted as the
finite temperature of the bulk theory. Thus the partition function
does indeed diverge as the geometry becomes degenerate. 

\bigskip

Given a generic $d$-dimensional metric which becomes degenerate in the
sense considered here as some parameter $u \rightarrow 0$ we can
construct a $(d+1)$-dimensional metric satisfying the equations
derived from (\ref{action}) as follows. Firstly, we should find the
conformal frame in which the $d$-dimensional metric determinant is
independent of $u$. Then we should write the higher-dimensional metric
in the form (\ref{cd}) and fix $\delta$ from the leading order terms
in the Einstein equations. $\gamma^2$ will follow from an expansion in
powers of $u$. 

\subsection{Interpretation of the bulk action}
\noindent

If one cannot remove all the divergences in the action with
covariantly defined counterterms, one has to decide how to interpret the bulk
action in terms of the dual conformal field theory. 
One suggestion - close in spirit to interpretations of the
Randall-Sundrum scenario in terms of the AdS/CFT correspondence
\cite{RS}, \cite{G}, \cite{W2} - is
the following. Instead of the ultimate goal being to take the
$\epsilon \rightarrow 0$ limit so that the cutoff boundary becomes the
true boundary, we need to keep the boundary at finite $\epsilon$. This
will ensure that the background geometry for the dual conformal field
theory is non-degenerate. 

In the Randall-Sundrum scenario \cite{RS}
an Einstein term is induced into the effective action
on the hypersurface, plus a cosmological term which we can effectively
adjust to zero by adding a brane tension term \cite{G}. The presence of
these two terms in the induced action is manifest from the counterterm
action (\ref{rsc}). However, for degenerate boundaries there is no
Einstein term in the ``hypersurface'' action. The leading order
propagator will follow from differentiating the 
action twice with respect to the hypersurface metric
$\tilde{\gamma}$. Unsurprisingly one can't get a
sensible brane world scenario from a higher-dimensional metric with
degenerate boundary.

The natural suggestion for the correspondence between the bulk and conformal
field theory partition functions is that we should simply take 
\be
I_{\rm{bulk}}(\epsilon) \approx  W_{\rm{cft}} (\epsilon), \label{hyp}
\ee
where $W_{\rm{cft}}(\epsilon)$ is the partition function 
for the conformal field theory in a geometry regulated by $\epsilon$.
Following the discussion in the last subsection, one
might question why we don't take the correspondence to be instead
\be
I_{\rm{nl}}(\epsilon) \sim W_{\rm{cft}} (\epsilon),
\ee
where we have removed the leading order divergence of the bulk action
by subtracting a counterterm of the form (\ref{surf}). 
However subtraction of such a counterterm would not be satisfactory
from a holographic point of view, since one would need to know the
index $\delta$ to carry out the subtraction {\it but} $\delta$ is not
known by the conformal boundary geometry. Another way of saying this
is that one effectively has to adjust the subtraction to the bulk geometry
rather than taking a generic subtraction. Of course in the
non-degenerate case one still needs to know that $\delta = 2$ to carry out 
the subtraction but the regularity of the geometry of the
regularisation limit implicitly tells us that $\delta = 2$. 

This proposal for the correspondence is equivalent to taking a strong
version of the holographic principle \cite{Su}: it assumes that
quantum gravity on any volume contained within a manifold 
can be described by a theory defined on the boundary of the volume.  
This is the basis for the recent work of \cite{G}, \cite{BK2},
\cite{V} but our proposal extends this
principle to more general negative curvature manifolds. 
One should be able to make more precise the correspondence between the
bulk field equations and the renormalisation group flow equations in
the conformal field theory along the lines of \cite{BK2}. The
difference will be that, in addition to the renormalisation group flow in
the conformal field theory as one flows in from infinity, one will also have
a flow in the effective target space geometry for the conformal field
theory. 

We should also mention that, although this procedure for cutting off
the interior path integral at a finite boundary seems to be the right
thing to do to compare partition functions, we probably need to be
more careful about how we do this. Simply cutting off the
theory will throw out some physics since it will not tell us about
physical processes in which particles propagate across our cutoff
boundary. However, our naive approach will be adequate for the
discussions here. 

To derive other thermodynamic quantities in the boundary conformal
field theory from the bulk, one would need to use the quasilocal 
tensor defined by Brown and York as \cite{BY}
\be
T^{\mu \nu} = \frac{2}{\sqrt{\tilde{\gamma}}} \frac{\delta
  I_{\rm{bulk}}}{\delta \tilde{\gamma}_{\mu \nu}},
\ee
and then define conserved quantities associated with Killing vectors
$\xi$ as
\be
Q_{\xi}(\epsilon) 
= \int_{\Sigma_{\epsilon}} d^2x \sqrt{\sigma} T_{\mu \nu} u^{\mu} \xi^{\nu},
\ee
where $u$ is the unit normal to a hypersurface
$\Sigma_{\epsilon}$ in $N_{\epsilon}$. The thermodynamic relation
between these quantities would be defined as usual as 
\be
I_{\rm{bulk}}(\epsilon) = \beta M(\epsilon) + ... - S(\epsilon),
\ee
where $\beta$ is the inverse temperature, and 
$M(\epsilon)$ and $S(\epsilon)$ correspond to the 
mass and entropy respectively of the regulated conformal field theory.

\subsection{Example 1: The Bergman metric}
\noindent

To check whether the bulk/boundary partition functions do diverge in
the same way, let us try to calculate both for some of the metrics 
discussed in \S\ref{two}. 
Suppose we introduce into the Bergman
metric an IR cutoff $\sinh \rho = l R \gg 1 $ so that the boundary
geometry is conformal to 
\be
ds^2 = l^2 R^2 (d\psi + \cos \theta d\phi)^2 + (d\theta^2 + \sin^2 \theta
d\phi^2); \label{geo1}
\ee
Then the bulk Euclidean action is
\be
I = - \frac{5\pi l^2 R^4}{4 G_{4}} - \frac{3 \pi R^2}{2 G_{4}}.
\ee
To calculate the surface term in (\ref{surf}), we need to bring the
metric near the conformal boundary into the form (\ref{cd}). Defining 
$x = 2 e^{-\sqrt{2} \rho}$ then the leading order terms in
the metric are
\be
ds^2 = \frac{dx^2}{l^2 x^2} + \frac{x^{-\frac{4 \sqrt{2}}{3}}} {2 l^2}
  \lbrace x^{-\frac{2 \sqrt{2}}{3}} 
(d \psi + \cos \theta d\phi)^2 + x^{ \frac{\sqrt{2}}{3}} (d\theta^2 +
\sin^2 \theta d\phi^2) \rbrace,
\ee
from which we see that we must take $\delta = 4 \sqrt{2} /3$ in
(\ref{cd}), and hence the first term in (\ref{surf}) becomes
\be
I_{\rm{surf}} = - \frac{5 \pi l^2 R^4}{4 G_4} + ..., 
\ee
which as expected coincides with the leading order divergence of the
effective action. 

\bigskip

As usual, strong coupling prevents us from calculating 
the partition function for the
associated conformal theory on the squashed three sphere directly;
however, in this case, we can calculate the $R$ dependence 
by an indirect method.
Supergravity in negative curvature Taub-Nut and Taub-Bolt
manifolds \cite{JM} also corresponds to the $(2 + 1)$ dimensional ``exotic''
conformal field theory \cite{SS}
which lives on the world volume of M2-branes
after placing them on a squashed three sphere. 
There is of course a very close relationship between the
AdS Taub-Bolt manifolds and the Bergman metric. The
Bergman metric is a radial extension of the second power of the Hopf bundle
over $S^2$ whilst the nut and bolt metrics are radial extensions of
the first power of the Hopf bundle over $S^2$ \cite{PP}. 
There is no problem in calculating the regularised Euclidean action
for the Taub-Nut and Taub-Bolt manifolds which have non-degenerate
boundaries. The metric for the nut solution {\footnote{The bolt
    solution does not exist in the parameter range relevant here.}} is 
\ba
ds^2 &=& V(r) (d \tau + 2 n \cos \theta d\phi)^2 + V^{-1}(r) dr^2 +
(r^2 - n^2) (d\theta^2 + \sin^2 \theta d\phi^2); \nonumber \\
V(r) &=& \frac{(r-n) (l^{2} r^2 + 2 n l^{2} r + 1 - 3 n^2 l^{2})}
{(r+n)},
\ea
and the action was calculated using counterterm subtraction 
in \cite{EJM} 
\be
I = \frac{4 \pi n^2}{G_{4}}(1 - 2n^2 l^2),
\ee
with the boundary geometry behaving as 
\be
ds^2 = 4 n^2 l^{2} 
(d\psi + \cos \theta d\phi)^2 + (d\theta^2 + \sin^2 \theta
d\phi^2), \label{geo2} 
\ee
where we identify $\tau \equiv \psi n$. 
The usual dictionary for the AdS/CFT correspondence \cite{Ma}
implies that we should take
\be 
N^{\frac{3}{2}} \approx \frac{1}{l^2 G_4},
\ee 
where $N$ is a measure of the number of unconfined degrees of freedom
for the gauge theory describing the dynamics of $N$ parallel M2-branes
wrapped on a squashed three sphere. 
So to compare the conformal field theory in the background geometry 
(\ref{geo2}) with that in (\ref{geo1}) we need to take the same values
of $l^2 G_4$ and set $R = 2 n $. In this limit the conformal
field theory partition function behaves as 
\be
I = \frac{\pi R^2}{G_4} (1 - \frac{R^2 l^2}{2}).
\ee
In the extreme squashing limit, the action diverges in the same way
as the bulk action for the Bergman metric. Of course, we shouldn't 
expect the coefficients to agree, since we can't assume that  
the two spacetimes 
correspond to the same state in the conformal field theory
{\footnote{Indeed, if we accept the hypothesis (\ref{hyp}) as true, 
    then the entropy for the Bergman metric is positive whereas that
    for the nut solutions is negative, so the Bergman metric
    corresponds to a highly excited state. 
    Of course the use of this argument is circular.
    Note that the negativity of the
    entropy for the nut solution can be viewed as a manifestation of
    the pathologies in the causal structure as discussed in \S2.}}. 
However,
since the degeneracy of the geometry will determine the leading order
divergence of the partition function, we should expect the actions to
diverge in the same way as we take $R$ to infinity. 

\bigskip

There is a possible flaw in the above argument.
It is not obvious that we can regulate the action for the
nut spacetime and then take a singular limit in $n$; 
these operations do not necessarily commute, since the spacetime  
becomes very singular as $n \rightarrow \infty$.
Although we should be reassured that
a very similar limiting process appears to work 
when one calculates the action for
critically rotating black holes \cite{HHT}, it would nice to check the
above conclusions in another way. 

Since the leading order behaviour of the partition function should not
depend on the details of the conformal field theory as $R \rightarrow
\infty$ it should be
reproduced by the partition function 
for free conformally coupled scalar and spinor fields in this
background. A related calculation was carried out in \cite{D}; 
it was found that if one considered
eigenmodes of a scalar field on a squashed sphere satisfying
\be
(-\nabla + \frac{1}{4}) \Phi_{k} = \lambda_k \Phi_k, \label{eqa}
\ee
then the partition function obtained from the zeta function $\zeta(s)
= \sum_{k} \lambda_{k}^{-s}$ did indeed
behave as $R^4$ in the extreme oblate limit. However,
this calculation is not directly relevant to conformally coupled
fields {\footnote{as Andy Strominger has also pointed out to me.}}: 
in the large $R$ limit, the operator (\ref{eqa}) is very
different from the conformally invariant operator 
\be
(-\nabla + \frac{R_g}{8}) = (-\nabla + \frac{1}{4} - \frac{R^2}{16}),
\label{cop}
\ee
where $R_g$ is the Ricci scalar. 
It is not difficult to apply the same techniques
to show that the divergence as $R^4$ persists for the
conformally coupled operator (\ref{eqa}); 
the analysis mirrors that of \cite{D} and is summarised in the Appendix. 
Note that there doesn't seem to be any natural intuitive explanation
for the $R^4$ dependence of the partition function; it follows
in a non-trivial way from the geometry. 

\subsection{Example 2: $H^2 \times H^2$}
\noindent

The second example we will consider 
is the product of two hyperbolic spaces $H^2 \times H^2$ whose
Einstein action is 
\be
I = - \frac{1}{48 \pi G_{4} l^2} \int \frac{d\theta dx_1 dx_2}{u^2 \sin^2
  \theta \cos^2 \theta} = - \frac{\sqrt{2} \sigma_3}{48 \pi G_{4} l^2 
u^2}, \label{bulk}
\ee 
where we have introduced a regulated volume
$\sigma_{3}$ for the volume of non-compact 
hypersurfaces of constant $u$ in the induced boundary metric 
\be
ds^2 = \frac{u^{\frac{4}{3}} d\theta^2}{2 \sin^2 \theta \cos^2 \theta}
+ \frac{dx_1^2}{u^{\frac{2}{3}} \cos^2 \theta} +
\frac{dx_2^2}{u^{\frac{4}{3}} \sin^2 \theta}. \label{non}
\ee
The metric can be brought into the form (\ref{cd}) 
with the choice of coordinate
\be
x = u^{\frac{\sqrt{2}}{\sqrt{3}}} \sin^{\frac{1}{\sqrt{6}}} \theta
\cos^{\frac{1}{\sqrt{6}}} \theta,
\ee 
and hence the above analysis is applicable here. 
Since there is in this case 
no obvious supergravity background with a related conformal boundary 
for which we can also calculate the action, the best that we can do is
to check whether we can reproduce this form of the partition
function from conformally coupled scalars in the background
(\ref{non}). The Ricci scalar for this metric is 
\be
R_g = -4 - 4 \cos^2 \theta \sin^2 \theta,
\ee
and modes of a conformally coupled 
scalar field behaving as $\phi(\theta,x_1,x_2)
\sim \phi_{\alpha}(\theta) e^{ik_i x_i}$ satisfy the equation
\ba
&&(2 \cos^2\theta \sin^2 \theta \partial_{\theta}^2 - \cos^2 \theta
k_1^2 u^2 - \sin^2 \theta k_2^2 u^2 + \frac{1}{2} +
\frac{1}{2} \cos^2 \theta \sin^2 \theta) \phi_{\alpha}(\theta) 
\\
&& \hspace{100mm} = -\lambda(\alpha, k_i u) \phi_{\alpha}(\theta). \nonumber
\ea
In fact we don't need to find the eigenvalues explicitly; all we need to know 
is that 
the eigenvalues $\lambda$ depend only on the combinations $(k_{i} u)$ and
$\alpha$. Furthermore, 
since the domain over which we are solving the equation is
non-compact, the index $\alpha$ is continuous and
the zeta function summation will take the form
\ba
\zeta(s) &=& \sum \lambda^{-s} =
\sigma_3 \int d\alpha dk_{i} \lambda(\alpha, k_{i}
u)^{-s}; \nonumber \\
&=& \sigma_{3} u^{-2} \tilde{\zeta}(s),
\ea
where $\sigma_3$ is again the regulated volume and 
in the latter equality $\tilde{\zeta}(s)$ is a function only of $s$.
Since the partition function can depend only on ${\zeta}'(0)$, it
manifestly exhibits the same behaviour as the bulk action
(\ref{bulk}), in agreement with our suggestion for the interpretation
of the bulk action.  
Suppose we interpret $x_1$ as the Euclidean time direction; then the
thermodynamic relation is given by
\be
I_{\rm{bulk}}(u) = \beta_{x_1} M(u),
\ee
where $\beta_{x_1}$ is the inverse temperature and the cutoff mass
$M(u)$ is negative. The entropy vanishes, which implies that $H^2
\times H^2$ corresponds in some sense to the ground state of the
conformal field theory, but the energy is negative which we should
probably interpret as discussed in \cite{E}.  

\section{Correlation functions in the boundary CFT} \label{four}
\noindent

In the previous section we considered how the bulk action
corresponds to the partition function for the conformal field
theory. The next question to ask is how the bulk supergravity action
acts as a generating functional for the correlation functions of the
conformal field theory. The analysis for the Bergman metric was
carried out in \cite{BSV}; the $su(2,1)$ symmetry of the bulk
corresponds to a $su(2,1)$ conformal symmetry group of the
boundary. This conformal symmetry is enough to fix the functional form
of two-point functions of scalar operators entirely, and 
this form is reproduced from the action for bulk scalar
fields. In particular, as in other cases of the AdS/CFT
correspondence, fields of a particular mass $m$ and spin $s$ are found to
correspond to scalar operators of definite 
conformal weights $\Delta(m,s)$ in the boundary CFT. 

Manifolds of degenerate boundary which fall into the first category of
\S\ref{two} can
hence be dealt with in much the same way as in the usual AdS/CFT
correspondence. However, the analysis is different for spaces
falling into the second category.  These spaces are characterised
by the existence of more than one infinite direction, not
linked by the symmetry group.  As we will discuss in this section, 
this means that massive scalar fields will give rise to boundary
data which is a sum of data of different conformal weights; the
relationship between the mass and the conformal weight in the CFT is
more subtle. A secondary characteristic of these spaces is that the
conformal symmetry group is not large enough to fix the form of
even the two point functions completely. 

\subsection{Two point functions from conformal symmetry}
\noindent

We will consider here the simplest non-trivial example, $H^2 \times
H^2$. Since the symmetry group of the manifold
is $sl(2,R) \times sl(2,R)$, which has a maximal compact subgroup of
$so(2) \times so(2)$, the boundary has only the latter group of symmetries. 
Expressed in terms of the $(u, \theta)$ coordinates, the Killing vectors
in the bulk are
\ba
k_1 &=& \partial_{x_1}; \hspace{5mm} k_2 = \partial_{x_2}; \\
l_1 &=& x_1 \partial_{x_1} + u \cos^2 \theta \partial_u - \cos \theta \sin
\theta \partial_\theta; \nonumber \\
l_2 &=& x_2 \partial_{x_2} + u \sin^2 \theta \partial_u + \cos \theta \sin
\theta \partial_\theta; \nonumber \\
m_1 &=& (x_1^2 - u^2 \cos^2 \theta) \partial_{x_1} + 2 x_1 u \cos^2
\theta \partial_u - 2 x_1 \cos \theta \sin \theta \partial_\theta;
\nonumber \\
m_2 &=& (x_2^2 - u^2 \sin^2 \theta) \partial_{x_2} + 2 x_2 u \sin^2
\theta \partial_u + 2 x_2 \cos \theta \sin \theta
\partial_\theta. \nonumber 
\ea
If one restricts to the boundary $u \rightarrow 0$, then the $k_i$
remain symmetries but the $l_i$ are conformal
symmetries only. Notice that one does not need the inverse metric to define
the conformal Killing vector equations and hence the conformal
symmetries are well defined even without a non-degenerate metric. 

Now let us consider how the two-point function of scalar fields $ \lbr
{\cal{O}}_{\Delta_1} (x) {\cal{O}}_{\Delta_2} (\bar{x}) \rbr$ is fixed by the
requirement of invariance under conformal transformations. Under a
conformal transformation generated by $\xi$ a field of conformal
weight $\Delta$ will transform as
\be
\delta_{\xi} {\cal O} = ({\cal{L}}_{\xi} + \frac{\Delta}{3} D_m \xi^m)
{\cal O},
\ee
where ${\cal L}$ is the Lie derivative. 
Then the
requirement of invariance under the isometries $k_i$ implies that the
two-point function only depends on the translationally invariant
quantities $(x_{1} - \bar{x}_{1})$ and $(x_2- \bar{x}_2)$. 
The requirement for the
two-point function to be covariant under the transformations generated
by the $l_i$ and $m_i$ is
\be
[ l_i^{(x)} + l_i^{(y)} ] \lbr {\cal{O}}_{\Delta_1}(x)
{\cal{O}}_{\Delta_2}(\bar{x}) \rbr 
= - \frac{1}{3} [ \Delta_1 D_{m}l_i^{m (x)} +
\Delta_2 D_m l_i^{m (y)} ] \lbr {\cal{O}}_{\Delta_1}(x)
{\cal{O}}_{\Delta_2}(\bar{x}) \rbr.
\ee
Now in this equation we need the inverse metric to be finite in order to 
define the right-hand side. To do this we note that if we use the
conformally rescaled boundary metric (\ref{non}) discussed in \S \ref{three}
then the metric determinant is independent of $u$ and
\be
D_m l_i^{m (x)} = \sin^2 \theta \cos^2 \theta \partial_m (\sin^{-2}
\theta \cos^{-2} \theta l_i^m).
\ee
Note that both the non-degenerate measure and the metric itself
have conformal dimension of minus three. 
Using the four conformal covariance 
conditions we can constrain the two-point function to
be of the form
\be
\lbr {\cal{O}}_{\Delta_1}(x)
{\cal{O}}_{\Delta_2}(\bar{x}) \rbr = \int d\chi f(\chi, \Delta_1, \Delta_2)
\frac{\sin^{\frac{2\Delta_1}{3} -
  \frac{\chi}{2}} \theta \sin^{\frac{2 \Delta_2}{3} -
  \frac{\chi}{2}} \bar\theta
\cos^{\frac{\chi}{2}} \theta \cos^{\frac{\chi}{2}} \bar\theta }{ (x_1 -
\bar{x}_1)^{\chi} (x_2 - \bar{x}_2)^{\frac{2 \Delta_1}{3} + 
\frac{2 \Delta_2}{3} - \chi}}. \label{Q1}
\ee
As expected the conformal symmetry group is not large enough to fix the
form of the two-point function completely. The function $f$ is not
fixed by symmetry and furthermore conformal
invariance does not fix $\Delta_1 = \Delta_2$; fields of unequal
conformal weight are not excluded from having a non-zero correlation
function. 

\subsection{Scalar fields in the bulk}
\noindent

Now let us consider how this form for the two-point function is
reproduced by the bulk theory. 
One of the most interesting
differences between this bulk boundary correspondence and
the usual non-degenerate correspondence is that a bulk scalar 
field of mass $m$ does not correspond to a single operator of weight
$\Delta(m)$. Instead, the scalar field acts as a source for a set of
operators of weights which depend not only on $m$ but also on the
``mode'' of the scalar field.

One can easily understand how this arises by looking at 
explicit solutions of the field equation. The field 
equation for a free scalar field of mass $m$ is
\be
[z_1^2 (\partial_{z_1}^2 + \partial_{x_1}^2) + z_2^2 (\partial_{z_2}^2
+ \partial_{x_2}^2 ) - m^2] \Phi^{m} = 0,
\ee
and so modes of the field behave as 
\be
\Phi^{m} \sim (z_1 z_2)^{1/2} K_{\nu}(k_1 z_1) K_{\sqrt{m^2 - \nu^2}}(k_2
z_2) e^{i k_1 x_1 + i k_2 x_2}, \label{eq1}
\ee
where we have chosen Bessel functions such that modes are bounded
at the interior points $z_1, z_2 \rightarrow \infty$. 
The allowed values of $m$ are determined by considering 
the spectrum of supergravity on $S^7$; we find that (in the units used
here) $m^2 \ge - 3/8$. We then restrict the allowed values of $\nu$ to
${\rm{Re}}(\nu) > 0$ to enforce boundedness in the interior. 
Note that most modes will consist of 
decaying oscillations in one hyperbolic space and exponential decay 
in the other. 

We should briefly mention that since the space is supersymmetric 
there are no unstable fluctuations of the scalar field; the point
is that although modes may be normalisable on one space they cannot be
simultaneously normalisable on the other. If one considers eigenmodes of 
$\Phi{(m)}$ with eigenvalues $\lambda_k$ then modes of negative $\lambda_k$
are not normalisable. 

A more elegant way of expressing the above analysis is in terms of 
representation theory. Solutions of the wave
equation for a massive scalar field form a representation of $sl(2,R)
\times sl(2,R)$, which can be decomposed as products of
representations of $sl(2,R)$ and $sl(2,R)'$ with Casimirs proportional
to $\nu^2$ and $(m^2 - \nu^2)$ respectively. Suppose we then consider primary
fields satisfying $k_1 \Psi = k_2 \Psi = 0$, 
$l_1 \Psi = - h_1 \Psi$ and $l_2 \Psi = - h_2 \Psi $, which behave 
as
\be
\Psi \sim u^{-h_1 - h_2} \cos^{-h_1} \theta \sin^{-h_2} \theta.
\ee
The quadratic Casimir is  
\be
m^2 \Psi = (l_1^2 + l_2^2 - \lbrace k_1, m_1 \rbrace - \lbrace k_2,
m_2 \rbrace ) \Psi = 
[h_1(h_1 +1) + h_2(h_2+1)] \Psi. \label{cw}
\ee
The conformal weights with respect to the two $sl(2,R)$
conformal groups are thus related by the mass of the bulk scalar
field, but are not fixed; this is the origin of the
$\chi$ integration in (\ref{Q1}). For fields of arbitrary spin $s$
the mass and conformal weight relation (\ref{cw}) becomes
\be
m^2 = [h_1(h_1 +1) + h_2(h_2+1)] - \frac{s^2}{2}.
\ee
Rewriting the scalar field in terms of the 
$(u,\theta)$ variables and taking the limit $u \rightarrow 0$ we get
\be 
\Phi^{m} \rightarrow u^{1 - \nu - \sqrt{m^2 - \nu^2}} (\cos^{1/2 -
  \nu} \theta \sin^{1/2 - \sqrt{m^2 - \nu^2}} \theta k_1^{-\nu}
k_2^{-\sqrt{m^2 - \nu^2}} e^{i k_1 x_1 + i k_2 x_2} ).
\ee
Note that the $u$ dependence is in general complex depending on 
the value of $\nu$. 
Explicitly the conformal weights $h_{i}$ are given in these variables
by
\be
h_1 = \nu - \frac{1}{2}; \hspace{5mm} h_2 = \sqrt{m^2 - \nu^2} -
\frac{1}{2}.
\ee
The total conformal weight of the boundary data will be determined by the $u$
dependence and is not independent of $\nu$; hence different modes of a
massive field will give rise to boundary data of different conformal
weight. 

\bigskip

The action for a free massive scalar field
reduces to the boundary term:
\be
I^{(m)} = \int dx dy \frac{d\theta}{u \sin^2 \theta \cos^2 \theta} \Phi^{m}
\partial_u \Phi^{m},
\ee
where in this equation and all that follow 
we are suppressing constant factors. Fourier transforming
(\ref{eq1}), the massive scalar field can be written in terms of
propagators on each hyperbolic space as
\be
\Phi^{(m)} = \int d\nu d\bar{x}_1 d\bar{x}_2 u^{\alpha + \beta} 
 \frac{\cos^{\alpha} \theta}{(u^2 \cos^2 \theta +
  (\Delta x_1)^2)^{\alpha}} \frac{\sin^{\beta} \theta}{(u^2 \sin^2 \theta +
  (\Delta x_2)^2)^{\beta}} \Phi^{(m)}(\nu,\bar{x}_1,
\bar{x}_2),
\ee
where $\alpha = \frac{1}{2} + \nu$ and $\beta = \frac{1}{2} + 
\sqrt{m^2 - \nu^2}$. 
In the limit  $u \rightarrow 0$, 
\be
\Phi^{(m)} \rightarrow  
\int d\nu u^{1 - \nu - \sqrt{m^2 -\nu^2}} \cos^{\alpha} 
\theta \sin^{\beta} \theta \Phi^{(m)}
(\nu,x_1,x_2). \label{eq5}
\ee
In these expressions we are drawing on the by now well known
propagators first discussed in \cite{W1}.
Note that we have not corrected the normalisation of the 
propagators following \cite{FMMR}
to ensure the right coefficients as $u \rightarrow 0$; in this expression, and 
all that follow, we will ignore $\nu$ dependent normalisation factors. 
We should allow the $\nu$ integration to run over all possible values. 
Furthermore, 
\ba
\partial_{u} \Phi^{(m)} & \rightarrow & \int d\nu d\bar{x}_1 d\bar{x}_2 
[ \frac{(\alpha + \beta) u^{\alpha + \beta -1}}{
  (\Delta x_1)^{2 \alpha}(\Delta x_2)^{2 \beta}}  
  - \frac{2 \alpha u^{1+ \alpha + \beta} \cos^2\theta}{(\Delta
  x_1)^{2(\alpha +1)}(\Delta x_2)^{2 \beta}} \label{eq6} \\
&& \hspace{30mm} - \frac{2 \beta u^{1+ \alpha + \beta} \sin^2\theta}{(\Delta
  x_1)^{2\alpha}(\Delta x_2)^{2(\beta+1)}}]  
\cos^{\alpha} \theta \sin^{\beta} \theta \Phi^{(m)} (\nu,x_1,x_2) \nonumber
\ea
Since ${\rm{Re}}(\alpha) > 0$ and ${\rm{Re}}(\beta) > 0$,
the first term is of leading order as $u \rightarrow 0$. However, as we 
shall see below, we cannot neglect the subleading terms in this case,
since these will give finite contributions to two point functions 
even as $u \rightarrow 0$. 

It is convenient at this stage to rewrite the last two integrals as
integrals over conformal weight of the boundary data, since we will
eventually want to compare predictions for two point functions with 
the boundary theory expectations. Then
\be
\Phi^{(m)}(u, \theta, x_1, x_2) \rightarrow \int d\lambda u^{-2\lambda/3}
Y^{(m)}(\lambda,\theta) \Phi^{(m)}(\lambda,x_1,x_2), \label{con2} 
\ee
where as we will see $-\lambda$ is the conformal weight of the
boundary data and we introduce the ``eigenfunctions''
\be
Y^{(m)}(\lambda,\theta) = \cos^{\alpha} \theta \sin^{\beta} \theta,
\ee
where $(\alpha + \beta) = \frac{2 \lambda}{3} + 2$ and in addition
\be
(\alpha - \frac{1}{2})^2 = m^2 - (\beta - \frac{1}{2})^2. \label{con1} 
\ee
It is also helpful to introduce the notation 
\be
K^{(m)}(\alpha, \beta, \theta, \Delta x_1, \Delta x_2) =
\frac{\cos^{\alpha} \theta \sin^{\beta} \theta}{(\Delta x_1)^{2
    \alpha} (\Delta x_2)^{2 \beta}},
\ee
and to simplify notation further we will suppress coordinate dependence
where obvious from now on. Then,
\ba
\partial_{u} \Phi^{(m)} & \rightarrow  & \int d\lambda d\bar{x}_1 d\bar{x}_2 
[ (2 + \frac{2 \lambda}{3}) u^{1 + 2\lambda/3}
K^{(m)}(\alpha, \beta) - 2 \alpha u^{3 + \frac{4 \lambda}{3}}
K^{(m)}((\alpha +1), \beta) \cos \theta \nonumber \\
 && \hspace{30mm} 
- 2 \beta u^{3 + \frac{4 \lambda}{3}} K^{(m)}(\alpha,(\beta+1)) \sin \theta]
\Phi^{(m)}(\lambda,\bar{x}_1,\bar{x}_2).
\ea
This action is of the form
\ba
I^{(m)}  &=& \int d\sigma
d\lambda d\bar{x}_1 d\bar{x}_2 d \bar{\lambda} u^{2 \bar{\lambda}/3
-2 \lambda/3} Y^{(m)}(\lambda,\theta) \Phi^{(m)}(\lambda)  
\lbrace (2 + \frac{2 \bar{\lambda}}{3}) 
K^{(m)}(\bar{\alpha},\bar{\beta}) \nonumber 
\\ && \hspace{20mm}
- 2 u^2 \left (\bar{\alpha} K^{(m)}(\bar{\alpha}+1,\bar{\beta})
 + \bar{\beta} K^{(m)}(\bar{\alpha},(\bar{\beta} +1)) \right ) \rbrace
\Phi^{(m)}(\bar{\lambda}), \label{eq7}
\ea
where $d\sigma$ is the non-degenerate measure on the boundary and 
$\alpha, \beta$ satisfy the constraints 
$(\bar{\alpha} + \bar{\beta}) = 2 + \frac{2 \bar{\lambda}}{3}$ as well as 
the constraint (\ref{con1}). 

\bigskip

The bulk/boundary correspondence tells us that the bulk scalar field
acts as a source for scalar operators in the boundary theory. 
One should hence associate the bulk action with terms in the
conformal field theory action of the form
\be 
I = \int d\lambda d\Delta d\sigma u^{-2 + \frac{2}{3} \Delta -
  \frac{2}{3} \lambda}
\Phi(\lambda,x) {\cal{O}}_{\Delta} (x), \label{eq4}
\ee
where $- \lambda$ is
the conformal weight of the boundary scalar field data and $\Delta$ is
the conformal weight of the operator $\cal{O}$. The $u$ dependence of
this action 
is determined by the requirement of conformal invariance:
suppose that the scalar field data behaves as 
\be
\Phi(u,x) \sim u^{-\lambda} \Phi^{b}(x),
\ee
on the boundary. That is, the data scales with $u$ which is a 
positive function that has a simple zero on the boundary. 
Following the same arguments as in \cite{W1},
the definition of $\Phi^{b}(x)$ depends on our particular choice of function, 
and if we transform $u \rightarrow e^{w} u$ then $\Phi^{b}(x)
\rightarrow e^{w \lambda} \Phi^{b}(x)$. Under the same transformation the
measure transforms as $d\sigma \rightarrow e^{2w} d\sigma$, since the
measure is of conformal weight $-3$. The
degeneracy of the boundary implies that an operator of conformal
weight $\Delta$ scales as $e^{-2 w \Delta/3}$ and since the action must be
conformally invariant this implies that it must be of the form 
(\ref{eq4}). 

\bigskip

Comparing the forms of (\ref{eq4}) and (\ref{eq7}) we see that we must
identify
\ba
{\cal{O}}_{\Delta}(x) &=& \int d\bar{x}_1 d \bar{x}_2 [ (\frac{2}{3}
\Delta) K^{(m)}(\alpha_1, \beta_1) \Phi^{(m)}(\Delta -3) \label{q1}
\\ && - (2 \alpha_2 K^{(m)}(\alpha_2+1, \beta_2)\cos \theta + 2 \beta_2
K^{(m)}(\alpha_2, \beta_2 +1)\sin\theta) \Phi^{(m)}(\Delta - 6)], \nonumber
\ea
where
\be
\alpha_1 + \beta_1 = \frac{2 \Delta}{3}; \hspace{5mm} \alpha_2 +
\beta_2 = \frac{2 \Delta}{3} - 2 .
\ee 
In addition $\alpha_i, \beta_i$ satisfy the constraint (\ref{con1}).
This gives us the expectation
value of the operator and functionally differentiating
this expression again will give us the two-point functions. 
Writing the boundary value of the scalar field as the mode expansion 
(\ref{con2}), then this expression may be inverted as 
\be
\Phi^{(m)}(\lambda,x_1,x_2) = u^{\frac{2\lambda}{3}} 
\int \frac{d \theta}{\sin^2
  \theta \cos^2 \theta} Y^{(m)}(\lambda,\theta) \Phi^{m}(u,x),
\ee
where we are again ignoring $\lambda$ dependent normalisation factors;
in fact, to get the normalisation factors right, we would have to
say more carefully how we are going to regularise these formally divergent
integrals. 
The two-point function $ \lbr {\cal{O}}_{\Delta}(x) {\cal{O}}_{\Delta'}
(y) \rbr$ is given by functionally differentiating (\ref{q1}) with respect to 
\be
u^{-2 + \frac{2 \Delta'}{3}} \Phi^{(m)}(u,x).
\ee
Let us consider the result of differentiating the first term in (\ref{q1});
then the correlation function is only non-zero for
$\Delta = \Delta'$ and we get a two-point function
\be
\lbr {\cal{O}}_{\Delta}(x) {\cal{O}}_{\Delta} (\bar{x}) 
\rbr = \frac{2 \Delta}{3} \frac{\cos^{\alpha_1} \theta \cos^{\alpha_1}
  \bar{\theta} \sin^{\beta_1} \theta \sin^{\beta_1} \bar{\theta}}{(\Delta
  x_1)^{2 \alpha_1}(\Delta x_2)^{2 \beta_1}}. \label{q2}
\ee 
(\ref{q2}) gives the leading order behaviour of the correlation
function for operators of the same conformal weight 
as $u \rightarrow 0$. Subleading behaviour is derived from
the last two terms in (\ref{q1}) 
\ba
\lbr {\cal{O}}_{\Delta}(x) {\cal{O}}_{\Delta} (\bar{x}) \rbr &=& - u^2 \lbrace 
2 \alpha_2 \frac{\cos^{\alpha_2 + 2} \theta \cos^{\alpha_2}
  \bar{\theta} \sin^{\beta_2} \theta \sin^{\beta_2} \bar{\theta}}{(\Delta
  x_1)^{2 (\alpha_2+1)}(\Delta x_2)^{2 \beta_2}} \nonumber \\
&& + 2 \beta_2 \frac{\cos^{\alpha_2} \theta \cos^{\alpha_2}
  \bar{\theta} \sin^{\beta_2+2} \theta \sin^{\beta_2} \bar{\theta}}{(\Delta
  x_1)^{2 \alpha_2}(\Delta x_2)^{2 (\beta_2+1)}} \rbrace.
\ea 
However, differentiating the last two terms in (\ref{q1}) also gives a non-zero
leading order contribution to the correlation function
\ba
\lbr {\cal{O}}_{\Delta}(x) {\cal{O}}_{\Delta-3} (\bar{x}) \rbr &=& - \lbrace 
2 \alpha_2 \frac{\cos^{\alpha_2 + 2} \theta \cos^{\alpha_2}
  \bar{\theta} \sin^{\beta_2} \theta \sin^{\beta_2} \bar{\theta}}{(\Delta
  x_1)^{2 (\alpha_2+1)}(\Delta x_2)^{2 \beta_2}} \nonumber \\
&& + 2 \beta_2 \frac{\cos^{\alpha_2} \theta \cos^{\alpha_2}
  \bar{\theta} \sin^{\beta_2+2} \theta \sin^{\beta_2} \bar{\theta}}{(\Delta
  x_1)^{2 \alpha_2}(\Delta x_2)^{2 (\beta_2+1)}} \rbrace. \label{q3}
\ea 
That is, operators of unequal conformal weight have a non-vanishing two point 
function!
Now (\ref{q2}) and (\ref{q3}) have precisely the same form as 
terms in the integral (\ref{Q1}); one can 
check that the equivalent between indices in the two expressions. Thus
we have explicitly verified that the scalar two point functions 
are reproduced from the action for bulk scalar fields. 

\bigskip

For completeness, let us now sketch the principle features of the
correspondence for the coset space $SO(3,1)/SO(2)$ given in
(\ref{coset}). The isometries of
this space are generated by the $sl(2,R) \times sl(2,R) \times so(2)$
algebra
\be
l_{0}^i =  i \partial_{\phi_i}; \hspace{5mm}
l_{\pm1}^{i} = i e^{\pm \phi_i} ( \coth \rho_i \partial_{\phi_i} \mp i
-\partial_{\rho_{i}}); \hspace{5mm} k = i \partial_{\psi}. 
\ee
One can find primary fields satisfying 
\be
l_{0}^{i} \Psi = h^{i} \Psi; \hspace{5mm} k \Psi = h_k \Psi;
\hspace{5mm} l_1^{i} \Psi = 0,
\ee
such that the bulk mass is related to these conformal weights as
\be
m^2 = h_{i} ( h_{i} + 1) + h_{k}^2.
\ee
By analogy to the above analysis for $H^2 \times H^2$, 
if one considers the correlation function of
two operators of conformal weights $\Delta_1$ and $\Delta_2$ on the
hypersurface $u \rightarrow 0$ defined from (\ref{cobound}), then 
\be
\lbr {\cal{O}}_{\Delta_1}(x) {\cal{O}}_{\Delta_2}
(\bar{x}) \rbr = \sum_{n,l} f_{n,l}(\Delta_1, \Delta_2) e^{i n \Delta
  \psi} U_{l}(\theta, \bar{\theta}, \Delta\phi_i, \Delta_1, \Delta_2),
\label{bndy}
\ee 
where the coefficients $f_{n,l}$ are not fixed by the conformal symmetry
but conformal covariance requires that $U_l$ satisfies four equations of
the form 
\ba
i (\partial_{\phi_1/\phi_2} - i \sin \theta \cos \theta \partial_{\theta} +
e^{i \Delta \phi_1/(-\Delta \phi_2)} (
\partial_{\bar{\phi}_1/\bar{\phi}_2} 
- i \sin \bar{\theta}
\cos \bar{\theta} \partial_{\bar{\theta}}) U_l \nonumber \\
= - \frac{2}{3} (
\Delta_1 \sin^2 \theta +  \Delta_2 \sin^2 \bar{\theta}  e^{i \Delta
  \phi_1/ (-\Delta \bar{\phi}_2)})U_l; \\
i (\partial_{\phi_1/\phi_2} + i \sin \theta \cos \theta \partial_{\theta} +
e^{(-i \Delta \phi_1)/\Delta \phi_2} ( \partial_{\bar{\phi}_1/\bar{\phi}_2} 
+ i \sin \bar{\theta}
\cos \bar{\theta} \partial_{\bar{\theta}}) U_l \nonumber \\
= - \frac{2}{3} (
\Delta_1 \cos^2 \theta +  \Delta_2 \cos^2 \bar{\theta}  e^{(-i \Delta
  \phi_1)/ \Delta \bar{\phi}_2})U_l. \nonumber
\ea
Terms in the summation (\ref{bndy}) should then be reproduced by considering
the action for bulk massive scalar fields. 

\bigskip

It is interesting to note that although the bulk scalar field action
for $H^2 \times H^2$
contains terms which diverge as $u \rightarrow 0$ the two-point
functions are actually regular in this limit. The same behaviour was
found for the Bergman metric; in fact in this case the action for a massive
scalar is independent of the IR regularisation parameter \cite{BSV}. Thus, we
don't need to keep $u$ finite in the correlation functions. 

If we keep $u$ finite in the boundary field theory, the two point 
functions will still be fixed by the conformal symmetry group of a
constant $u$ hypersurface. To compare with the bulk theory, we should
introduce propagators describing sources at finite $u$ and repeat the
above analysis. It would be interesting to consider the flow of the
propagators as one changes $u$, particularly in the context of
making more precise the correspondence at finite $u$. One could also
compare the spectrum on $H^2 \times H^2$ with operators appearing 
in the boundary theory.

\acknowledgements

I would like to thank Finn Larsen for discussions in the early stages
of this work and to thank Harvard University, where
this work was begun, for hospitality. 
Financial support for this work was provided by St John's College,
Cambridge.

\appendix
\section*{Effective actions on the squashed three sphere}
\noindent
Eigenvalues of the scalar operator (\ref{eqa}) used in \cite{D} are
\be
\lambda = \frac{1}{4 l_{3}^2} (n^2 + 4 (l_3^2 -1)(q + \frac{1}{2})(n -
q - \frac{1}{2}),
\ee
with degeneracy $n = 1,\infty$. We have set $l_3^2 = l^2 R^2$ and $q$
runs from $0$ to $(n-1)$. For our conformally coupled operator
(\ref{cop}) we just
need to shift the eigenvalues, and can hence write the partition
function as $W_{\rm{sc}} = \frac{1}{2} \zeta'(0)$ where
\be
\zeta(s) = (2 l_3)^{2s} \sum_{n=1}^{\infty} \sum_{q=0}^{n-1}
\frac{n}{(n^2 + 4 (l_3^2 -1)(q + \frac{1}{2})(n - q - \frac{1}{2}) -
  \frac{l_3^4}{4})^{s}}.
\ee
The approach of \cite{D} was to apply the Plana summation formula to
the $q$ summation; in the extreme oblate limit $l_3 \rightarrow
\infty$ it is then quite easy to find the dominant term in the zeta function
which behaves as $l_3^4$. 

Following the same approach here, the key point is that, although
sub-dominant terms in the Plana summation formula are affected by the
shift in eigenvalues, the dominant term in the extreme prolate limit is
determined by a very similar term to that in \cite{D}
\be
\zeta(s) \approx 2 i (2 l_3)^{2s} \int_{0}^{\infty} \frac{dt}{exp(2
  \pi t) + 1} \lbrace \frac{n}{(n^2 + 4 (l_3^2 -1)(t^2 - i t n) -
  \frac{l_3^4}{4})^s} - ( t \rightarrow - t) \rbrace, 
\ee
which we can analyse by the Watson-Sommerfeld method to give a leading
order contribution of 
\be
W_{\rm{sc}} \approx \frac{3 l_3^4}{2 \pi^2} \zeta_{R}(3),
\ee
as was found in \cite{D}.

\end{document}